\documentclass[twocolumn,showpacs,preprintnumbers,amsmath,amssymb,epsfig,widetext]{revtex4}

\usepackage{graphicx}
\usepackage{dcolumn}
\usepackage{bm}
\usepackage{epsfig}

\def\fun#1#2{\lower3.6pt\vbox{\baselineskip0pt\lineskip.9pt
  \ialign{$\mathsurround=0pt#1\hfil##\hfil$\crcr#2\crcr\sim\crcr}}}

\input epsf

\def\plotone#1{\centering \leavevmode
\epsfxsize= 1.0\columnwidth \epsfbox{#1}}

\def\be{\begin{equation}}
\def\ee{\end{equation}}
\def\ba{\begin{eqnarray}}
\def\ea{\end{eqnarray}}
\def\half{\frac{1}{2}}
\def\nn{\nonumber}

\begin{document}

\preprint{}

\title{Constraints on supernovae dimming from photon-pseudo scalar coupling}

\author{Yong-Seon Song}
\email{ysong@cfcp.uchicago.edu}
\author{Wayne Hu}
\email{whu@background.uchicago.edu}
\affiliation{Kavli Institute for Cosmological Physics,
Department of Astronomy \& Astrophysics, and Enrico Fermi Institute, University of Chicago, Chicago IL 60637 
}

\date{\today}

\begin{abstract}
An alternative mechanism that dims high redshift 
supernovae without cosmic acceleration utilizes an oscillation of photons
into a pseudo-scalar particle during transit.  Since angular diameter distance 
measures are immune to the loss of photons,
this ambiguity in interpretation 
can be resolved by combining  CMB acoustic peak measurements with 
the recent baryon oscillation detection in galaxy power spectra.
This combination excludes a non-accelerating dark energy species at the
$4\sigma$ level 
regardless of the level of the pseudo-scalar coupling.  While solutions still exist with
substantial non-cosmological dimming of supernovae, they may be tested with 
future improvement in 
baryon oscillation experiments.
\end{abstract}

\pacs{draft}


\maketitle

\section{introduction}
Cosmic acceleration is mainly  inferred through geometrical measures
of the expansion history, most notably from the dimming of high redshift supernovae \cite{perlmutter98,perlmutter99,riess98,riess04}.
In order to be certain of this interpretation, all other plausible explanations
of the dimming must be ruled out.

An alternative mechanism has been proposed for the dimming of supernovae \cite{csaki01}.
Here photons are converted into an unseen particle during transit.  This mechanism 
is based on a generic interaction 
between the photon and a pseudo-scalar field. 
Other cosmological consequences of this pseudo-scalar (p-p) 
coupling include Faraday-like rotations, spontaneous generation of
polarization, and primordial magnetic fields generation 
\cite{carroll91,lue98,das04,bertolami05}.

 Mediated by an 
external magnetic field, this coupling  also produces the conversion that
dims supernovae.  This mechanism is similar to
that exploited by axion direct detection experiments  \cite{Sik83} but in the reverse
direction.  Previous works examining the viability of this alternative 
dimming mechanism have largely focused on constraints from the
observed achromaticity of the dimming \cite{deffayet01,mortsell02}.

In this Paper, we reconsider this alternate mechanism in light
of recent cosmological measurements. 
First, we show that  there exists large regions in the parameter space
that both satisfy current luminosity distance measurements and 
avoid achromaticity constraints  by saturating the conversion.
Hence exclusion of this alternate explanation solely through supernovae
data is difficult.   

On the other hand, cosmological probes of acceleration 
that do not involve luminosity distances are immune to dimming effects.
Specifically, expansion history constraints based on angular diameter
distances from the measurement of standard rulers
 can be used to directly test the dimming hypothesis.
The acoustic oscillations set by the photon-baryon fluid before
recombination imprints the sound horizon as a standard ruler in 
both the cosmic microwave background and galaxy power spectra.
We combine  measurements from WMAP \cite{spergel03} 
which give the angular
diameter distance to recombination with the recent detection of 
baryon oscillations in the SDSS LRG
galaxy survey \cite{eisenstein05} which provides the distance ratio
between the galaxies and the CMB.

Our result exclude a non-accelerating species of dark energy at the 4$\sigma$ level
even allowing for the possibility of photon conversion.
 Furthermore,
within the 95\% confidence limits there exists separate solutions
with no dimming and strong dimming of supernovae both of which require
cosmic acceleration.   These possibilities can be distinguished in the future with better
constraints from the acoustic oscillations.  

In an Appendix, we also
consider spectral constraints from COBE FIRAS
on the dimming of the CMB itself.   These constraints, which are more robust than and equally 
powerful compared to CMB anisotropy constraints \cite{csaki01},  are useful for
eliminating astrophysically disfavored regions of the parameter space that involve large external
magnetic fields with a short coherence length.

\section{photon-pseudo scalar conversion}

We study the interaction between photons and a pseudo-scalar field $\phi$ through
\ba\label{action}
S
&=&\int d^4x \sqrt{-g} 
\Big[\half\partial_{\mu}\phi\partial^{\mu}\phi-\half m^2\phi^2\nn \\
&&-\frac{1}{4}F_{\mu\nu}F^{\mu\nu}
+\frac{\phi}{4M} F_{\mu\nu}\tilde F^{\mu\nu} 
\Big] \,,
\ea
where $\tilde{F}^{\mu\nu}=\half\epsilon^{\mu\nu\alpha\beta}F_{\alpha\beta}$
and $\epsilon^{\mu\nu\alpha\beta}$ is the Levi-Civita tensor.

In a small patch where the effect of the cosmic expansion
is negligible, the equation of motion for $\phi$ is
\ba
\frac{\partial^2\phi}{\partial t^2}-\nabla^2\phi 
=-m^2\phi
-\frac{1}{M} \vec{E}\cdot\vec{B}.
\ea
The magnetic field $\vec{B}$ is a combination of the internal magnetic field
of the radiation and any external magnetic field that might exist.  
We assume however that within a domain the external magnetic
field dominates.  
The external magnetic field in the intergalactic medium
is expected to be around $10^{-10}{\rm\,\,Gauss}$ 
with 1Mpc comoving domain size \cite{furlanetto01,kronberg93}.

\begin{figure}[htbp]
  \begin{center}
    \plotone{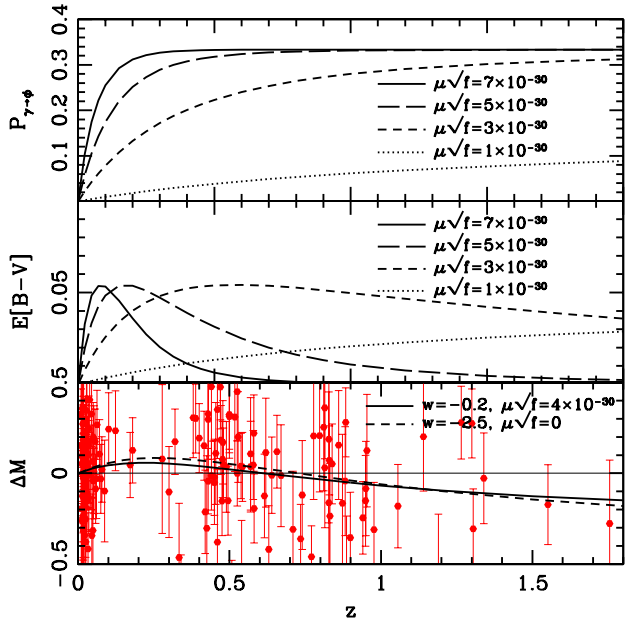}
    \caption{Top panel: the conversion probability with a given $\mu\sqrt{f}$ as a function of
    redshift $z$ (see text).
Middle panel: the magnitude change in the dimming between B and V bands.
Bottom panel: dimmed bolometric magnitude
of the supernovae in several cosmic acceleration scenarios parameterized
by the dark energy equation of state $w=p/\rho$ and 
the p-p coupling scenario.  Data is plotted relative to the
an accelerating $\Lambda$CDM model
with $w=-1$ (horizontal line).
}
\label{fig:chromaticdimming}
\end{center}
\end{figure}

We split the electric Maxwell equation into two components,
$E_{\|}$ parallel to $\vec{B}$ and $E_{\bot}$ perpendicular to $\vec{B}$.
The equation of motion for $E_{\|}$ is
\ba
\frac{\partial^2}{\partial t^2}E_{\|}-\nabla^2E_{\|}
=\frac{1}{M}\frac{\partial^2\phi}{\partial t^2}B_{\|}.
\ea
Here $E_{\|}$ is coupled to the scalar field by the external
magnetic field, but there is no mixture between $E_{\bot}$ and $\vec{B}$.
To summarize the coupled equations in matrix form,
\ba
 \left[ \frac{\partial^2}{\partial l^2}+\omega^2 \right]
 \left( \begin{array}{ccc} 
    A_{\|}\\ 
    A_{\bot}\\ 
    \phi\\
 \end{array}\right) 
=
 \left( \begin{array}{ccc}
    \omega_p^2 & 0 & \mu\omega\\
    0 & \omega_p^2 & 0\\
    \mu\omega & 0 & m^2\\
 \end{array}\right)
 \left( \begin{array}{ccc}
    A_{\|}\\
    A_{\bot}\\
    \phi\\
 \end{array}\right)\,,
\ea
where $\vec A = \vec E/\omega$, $\mu=B_{\|}/M$
and $l$ is the physical distance
traveled in each domain.

The off-diagonal interaction terms 
induce mixing between $A_{\|}$ and $\phi$. 
The eigenvalues and the eigenstates of
this mixing are given by
\ba
\lambda_{\mp}=\frac{\omega_p^2+m^2}{2}\mp\frac{1}{2}
\sqrt{(\omega_p^2-m^2)^2+4\mu^2\omega^2}
\ea
and
\ba
 \left( \begin{array}{cc}
    \lambda_-\\
    \lambda_+\\
 \end{array}\right)
=
 \left( \begin{array}{cc}
    \cos\theta & -\sin\theta\\
    \sin\theta & \cos\theta\\
 \end{array}\right)
 \left( \begin{array}{cc}
    A_{\|}\\
    \phi\\
 \end{array}\right)\,,
\ea
where the mixing angle is given by
\ba
\tan 2\theta=\frac{2\mu\omega}{\omega_p^2-m^2}.
\label{eqn:mixingangle}
\ea

The conversion probability of a photon polarized in the $\parallel$ direction 
in a single external magnetic domain
is 
\ba
P_{\gamma \rightarrow \phi} 
=\sin^2 2\theta
\sin^2\left[s\,
\frac{\sqrt{(\omega_p^2-m^2)^2+4\mu^2\omega^2}}{4\omega}\right]\,,
\label{eqn:conversionrate}
\ea
where $s$ is the physical size of a coherent field domain
and $\omega_p$ is the plasma frequency.  As pointed out by \cite{deffayet01}, the
finite plasma frequency of the intergalactic medium sets a lower limit on the
effective mass scale.  We will hereafter assume that $\omega_p \gg m$ throughout.

\section{simulating dimming}
We now simulate the dimming of supernovae through multiple field domains in 
a cosmological context.  We assume passive evolution of the parameters
such that the field strength scales
with the scale factor $a$ as
$B \propto a^2$,
 the domain size
is fixed in comoving coordinates $s \propto a$, and
 the plasma frequency 
\begin{equation}
\omega_p = 1.2\times 10^{-14}\left( { n_{e0} \over 10^{-7} {\rm cm^{-3} } }\right)^{1/2}  a^{-3/2} {\rm eV}\,, 
\end{equation}
where $n_{e0}$ is the free electron density today.  

A relevant parameter to consider is the ratio of the coherence length $s$ to the plasma 
length $l_p \equiv 2 \omega/\omega_p^2$ at the observation frequency \cite{deffayet01}.  If this 
ratio is large, then the  spatially oscillating piece in the conversion probability
 Eqn.~(\ref{eqn:conversionrate})
will be averaged to $1/2$ across the many random domains
such that
\ba
P_{\gamma \rightarrow \phi} 
\approx {1 \over 2}\sin^2 2\theta
\ea
in each domain.
For optical light at $\omega_{\rm SN} \sim 1$eV,
the plasma length today  $l_{p0} \sim 100 (n_{e0}/10^{-7} {\rm cm^{-3}})^{-1}$ kpc.
With a typical magnetic domain size of order a Mpc, we expect $s \gg l_p$ is a good assumption.
 Since the mixing angle scales with the plasma frequency as given in
Eqn.~(\ref{eqn:mixingangle}), we will take a fiducial value of  $n_{e0}=2.4\times10^{-7}$ cm$^{-3}$
with the understanding that other cases may be obtained by replacing
\ba
\mu \rightarrow \mu \left( {n_{e0} \over 2.4\times10^{-7} {\rm cm}^{-3} }\right)^{-1}\,.
\ea
Note that in this limit, the conversion probability  in a single domain is highly chromatic
with a scaling of $P_{\gamma \rightarrow \phi} \propto \omega^2$ \cite{deffayet01}.

As an aside,
there is also empirical evidence disfavoring dimming in the opposite limit of 
 $s \ll l_p$.  Here $P_{\gamma \rightarrow \phi} \approx 4 s^2\mu^2$
and becomes independent of frequency. Hence low frequency photons from cosmological
sources can be substantially dimmed.   The blackbody nature of the CMB places
significant constraints on dimming in this regime.   
In the Appendix we show that 
\begin{equation}
P_{\gamma \rightarrow \phi}(\omega_{\rm CMB})< 2.5 \times 10^{-4} \quad (95\%{\rm CL})\,,
\end{equation}
where $\omega_{\rm CMB} =1.24 \times 10^{-3}{\rm eV}$. 
If at $\omega_{\rm CMB}$ the coherence length is still $s \ll l_p$ then substantial
dimming of supernovae would be completely ruled out.

 Under the assumption that the transition $s \sim l_p$ lies
somewhere between the optical and CMB frequencies, this constraint translates to
one on  $l_{p0}/s$ through
\ba
 { P_{\gamma \rightarrow \phi}(\omega_{\rm CMB}) \over P_{\gamma \rightarrow \phi}(\omega_{\rm SN})}
\approx 2\left(\frac{l_{p0}}{s} {\omega_{\rm CMB} \over \omega_{\rm SN}} \right)^2
\frac{D(z_{\rm CMB})}{D(z_{\rm SN})}
a_{\rm SN}^{2},
\ea
where $D(z)$ is the comoving distance to redshift $z$, $l_{p0}$ is evaluated at
the SN frequency, and $z_{\rm CMB}$ is the highest redshift for which the magnetic field
persists.   Note that this $10^{-4}$ 
constraint from spectral distortions is more robust than and equally powerful
compared with constraints based on the angular variation of the dimming.   
Although the latter lie at an amplitude of $10^{-5}$, anisotropy measurements are
typically taken at a frequency that is up to an order of magnitude smaller (30GHz vs 300GHz).
The strong frequency scaling of $P_{\gamma \rightarrow \phi} \propto \omega^2$
more than compensates the lower precision of the spectral measurements. 
With supernovae at $z_{\rm SN} \approx 0.5$ dimmed by order unity
and assuming $z_{\rm CMB}$ is high, e.g.~comparable to the reionization redshift,
$s/l_p > 1/10$.  Hence
a very small coherence length to the field is not allowed if the supernovae are to
be dimmed by order unity.    We hereafter assume that $s \gg l_p$ for viable models.

To handle multiple domains, we employ a density matrix formulation.
The initial conditions at emission are given by unpolarized photons, arbitrarily normalized
to unity for convenience, and no excitation of the pseudo-scalar field
\ba
  S_0^{ab}(s_0) \equiv S_1^{ab}(0)= \left(
  \begin{array}{ccc}
    \half & 0  & 0 \\
    0 & \half  & 0   \\
    0 & 0  & 0  \\
  \end{array}\right)\,.
\ea
In our notation for the density matrix $S_i^{ab}(l)$,
 $a$ and $b$ denote the components $(\|,\bot,\phi)$ with respect to the magnetic field of the
domain, the subscript $i$ denotes the
$i$th magnetic field domain, and the argument $l$ denotes the physical location in the domain,
$0$ for the beginning, $s_i$ for the end.

\begin{figure*}[t]
\centerline{
\epsfxsize=3.3truein\epsffile{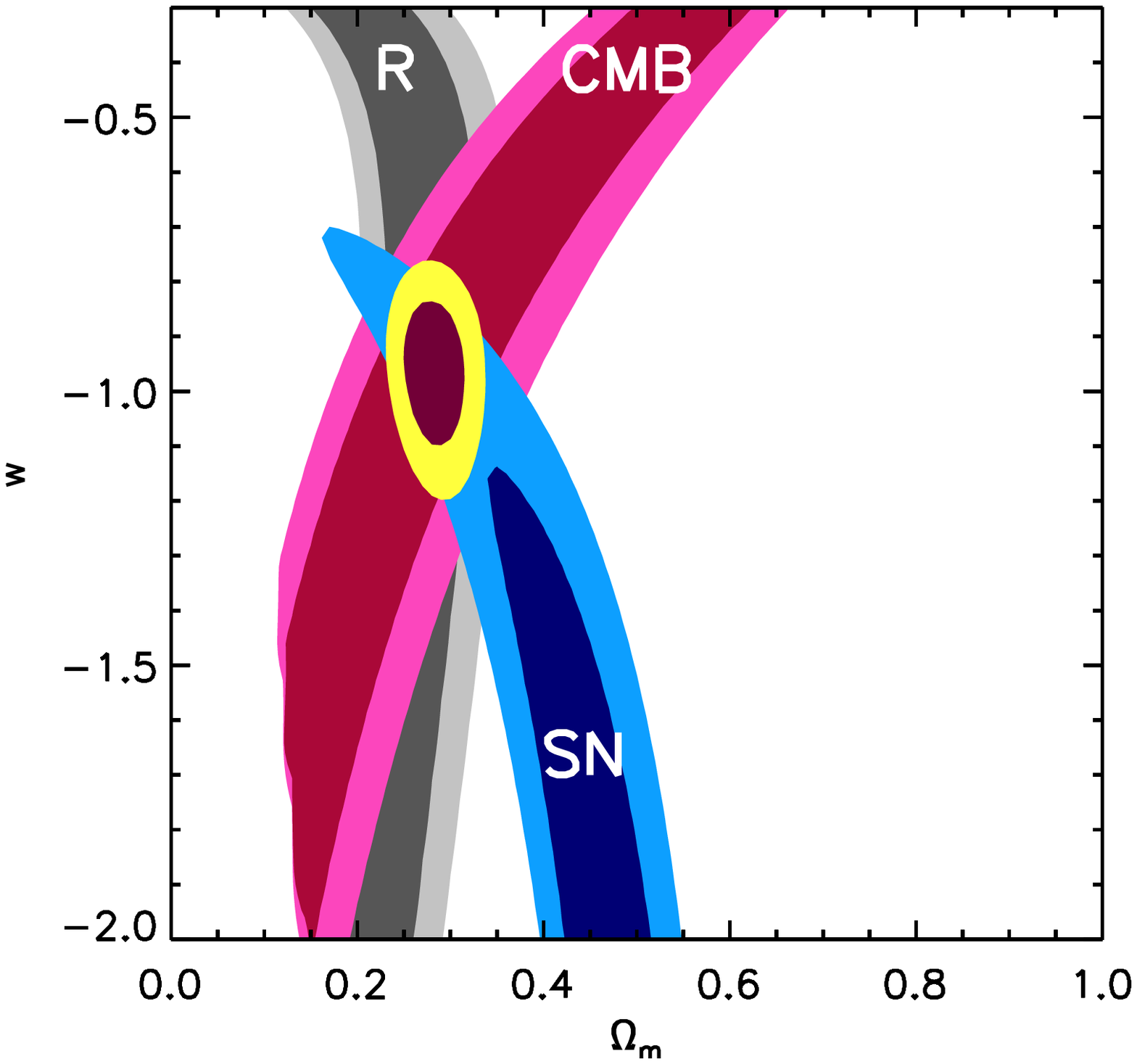}$\,$
\epsfxsize=3.3truein\epsffile{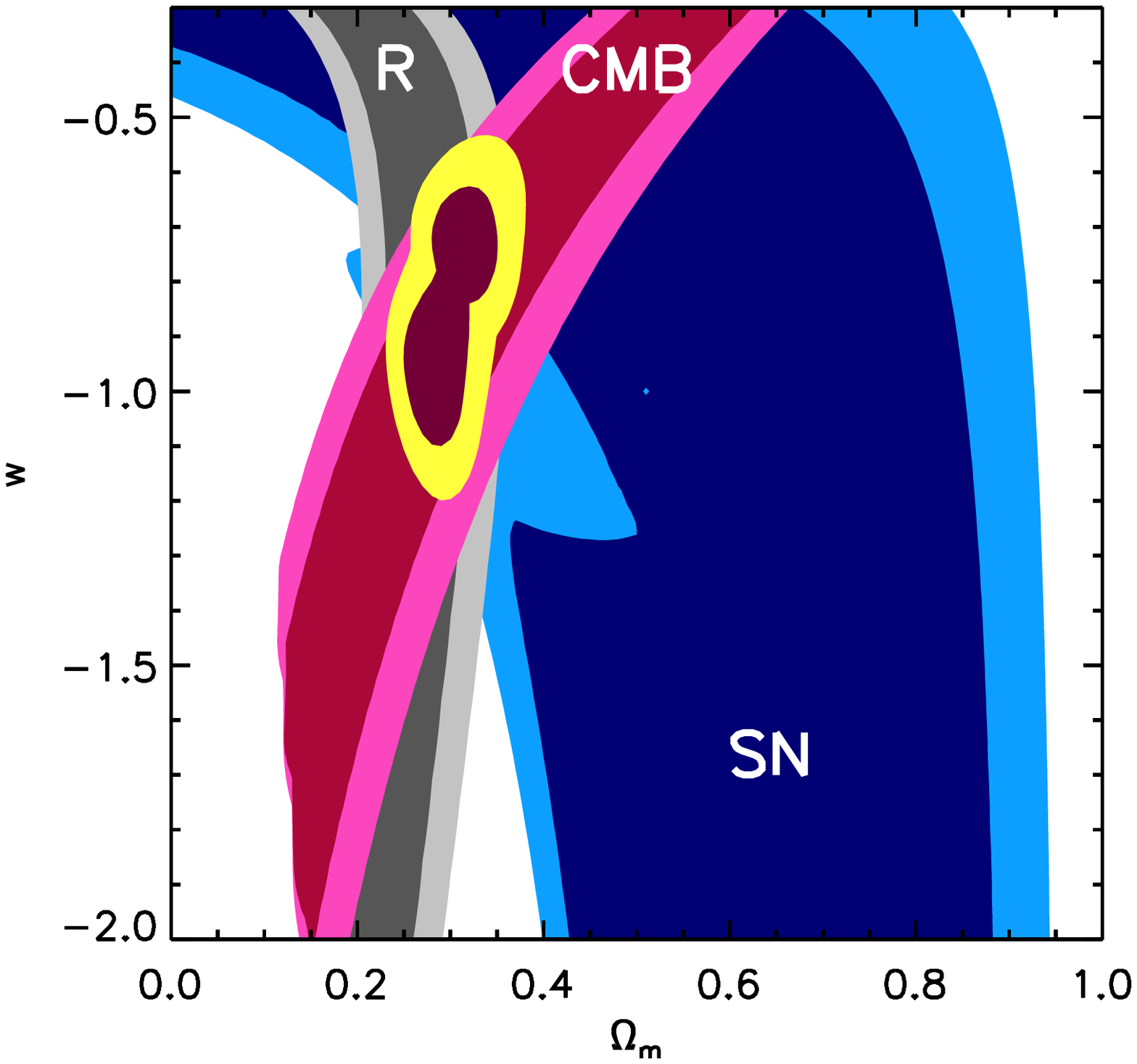}$\,$}
\caption{$\Omega_m-w$ constraints at the 68\% and 95\% CL. 
Cosmologies without p-p dimming are shown in the left panel and those with
dimming in the right.   Separate contours denote supernovae constraints (SN),
baryon oscillations (R), and CMB acoustic peak constraints (CMB) as labeled. The
ellipses at the intersection denote the combination of all constraints.
}
\label{fig:fitting}
\end{figure*}

Since the orientation of $\vec B$ in each domain is not identical, the density matrix must be
rotated between each domain.  
Defining $\xi$ as  the angle between $\vec B$ and
 the plane perpendicular to the propagation $B_{\|}=\cos\xi B$. With $\psi$ as the angle 
between $B_{\|}$ in two neighboring domains, the density matrix 
transforms as
\begin{widetext}
\ba
  S_i^{ab}(0) = 
\left(
  \begin{array}{ccc}
    \cos^2\psi S_{i-1}^{\|\|}+\sin^2\psi S_{i-1}^{\bot\bot}-\sin2\psi S_{i-1}^{\|\bot} & \half\sin2\psi(S_{i-1}^{\|\|}-S_{i-1}^{\bot\bot})+\cos2\psi S_{i-1}^{\|\bot}  & \cos\psi S_{i-1}^{\|\phi}-\sin\psi S_{i-1}^{\bot\phi} \\
    \half\sin2\psi(S_{i-1}^{\|\|}-S_{i-1}^{\bot\bot})+\cos2\psi S_{i-1}^{\|\bot} & \sin^2\psi S_{i-1}^{\|\|}+\cos^2\psi S_{i-1}^{\bot\bot}+\sin2\psi S_{i-1}^{\|\bot}  &   \sin\psi S_{i-1}^{\|\phi}+\cos\psi S_{i-1}^{\bot\phi} \\
    \cos\psi S_{i-1}^{\phi\|}-\sin\psi S_{i-1}^{\phi\bot} & \sin\psi S_{i-1}^{\phi\|}+\cos\psi S_{i-1}^{\phi\bot}  &  S_{i-1}^{\phi\phi} 
  \end{array}\right)\,,
\ea
where the right hand side is evaluated at position $s_i$ of the previous domain.  We chose random variates uniform over $2\pi$
for the angles in each domain.

To propagate the density matrix across $i$, we employ the conversion probabilities
implied by the mixing matrix
\ba
S_i^{\|\|}(s_i)&=&(1-\half\sin^22\theta)S_{i}^{\|\|}(0)
+\half\sin^22\theta S_{i}^{\phi\phi}(0)+\sin2\theta\cos2\theta S_{i}^{\|\phi}(0)\,, \nn \\
S_i^{\|\phi}(s_i)&=&\half\sin2\theta\cos2\theta S_{i}^{\|\|}(0)
-\half\sin2\theta\cos2\theta S_{i}^{\phi\phi}(0)+\sin^22\theta S_i^{\|\phi}(0)\,,  \nn \\
S_i^{\phi\phi}(s_i)&=&\half\sin^22\theta S_i^{\|\|}(0)+(1-\half\sin^22\theta)S_{i}^{\phi\phi}(0)-\sin2\theta\cos2\theta S_{i}^{\|\phi}(0)\,. \nn
\ea
There is no conversion in other components.
\end{widetext}

The total conversion probability at the observer is determined by the appearance
probability of the pseudo-scalar field at the last domain.
The total conversion probability  is given by
\ba
P_{\gamma\rightarrow\phi}=S_{i=D(z)f/S}^{\phi\phi}(s)\,,
\label{eqn:totalconversion}
\ea
where $f$ is the one dimensional fraction of 
the external magnetic domain between the observer
and the source at a comoving distance  $D(z)$.  Here $S = s_i/a_i$ is the
assumed constant coherence length in comoving coordinates.

As can be seen in Eqn.~(\ref{eqn:totalconversion}),  the total conversion
is a function of  $\mu\sqrt{f}$. 
In the top panel of Fig~\ref{fig:chromaticdimming}, we show the conversion 
as a function of $z$ for several choices of  $\mu\sqrt{f}$.
Given a  limit on the mass scale of the coupling of $M \ge 2\times 10^{19}{\rm eV}$
\cite{groom00,raffelt99}, the magnetic field strength must be of order
$10^{-10} {\rm Gauss}$ to establish a substantial conversion across a cosmological
distance. This is a plausible strength for the
intergalactic field.  

The bottom panel of Fig~\ref{fig:chromaticdimming} shows that
the dimming  produced by an accelerating component of dark energy
with an equation of state $w=p/\rho < -1/3$  and by the p-p coupling with $w>-1/3$
looks almost identical (see the following section for model details).  
We would expect, as shown explicitly in the next section, that
both scenarios would be equally favored by the luminosity
distance measures alone.

To mimic cosmic acceleration, the dimming must also be achromatic.
Although the conversion probability in an individual domain is highly chromatic,
the net dimming need not be so.
The conversion probability equilibrates at $1/3$ due to the oscillation of the pseudo-scalar
back to a photon.
Once the conversion probability is saturated, the dimming becomes achromatic.

With a sufficiently large $\mu \sqrt{f}$ achromaticity can be achieved at a sufficiently
 low redshift to avoid observational constraints.
Observations limit the magnitude difference of the dimming 
between B($0.44\mu m$) and V($0.55\mu m$) bands to be at most $0.03$ \cite{deffayet01}.
But as we see in the middle panel of Fig~\ref{fig:chromaticdimming},
once the conversion probability 
reaches the saturation limit, the difference in dimming between the 
bands disappears.

Thus in terms of supernovae data alone, dimming by the p-p coupling and cosmic 
acceleration are equally favored.

\section{model constraints}

We now explore the allowed parameter space of p-p dimming and cosmic acceleration
models.  
The cosmic acceleration is parametrized by the dark energy equation of state 
$w=p/\rho$ and the non-relativistic energy
density relative to critical $\Omega_m$ in a flat universe.  
The p-p dimming model is  parameterized by $\mu\sqrt{f}$. For simplicity, we will
fix $f=2/3$ and quote the allowed region in $\mu$ evaluated at the present epoch.

\begin{figure}[htbp]
  \begin{center}
    \plotone{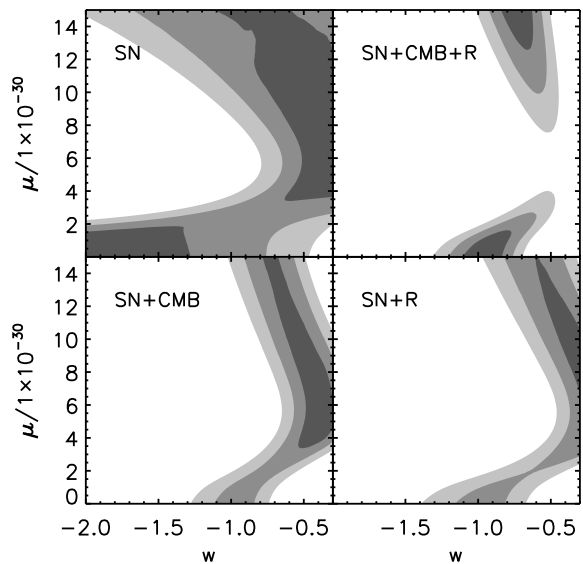}
    \caption{$w-\mu$ constraints at the 68\%, 95\% and 99\% CL. The 
    magnetic field filling fraction $f$ is fixed to 2/3 for simplicity.
}
\label{fig:mu_w}
\end{center}
\end{figure}

First, we use the gold set of the supernovae data alone \cite{riess04}.
When we consider the supernovae dimming just by the cosmic acceleration,
a phantom dark energy model with $w<-1$ is marginally  favored 
as shown in the left panel of Fig~\ref{fig:fitting} (see e.g. \cite{HutCoo05}).
It is interesting to note that 
the bolometric dimming of the supernovae most favored by
the cosmic acceleration scenario
is statistically indistinguishable from the best fit p-p coupling scenario with $w=-0.2$ 
and no acceleration,
as shown in the bottom panel of Fig~\ref{fig:chromaticdimming}.
This degeneracy is quantified in the top-left panel in Fig~\ref{fig:mu_w}. It shows that
the cosmic acceleration by the phantom dark energy model
with the negligible  p-p coupling
is almost equally favored compared with  strong p-p coupling
with no cosmic acceleration \cite{CsaKalTer05}.  The two appear as discrete solutions with the 
$\Lambda$CDM model in the weakly disfavored intermediate regime.

Other cosmological tests are required to break this degeneracy.
We next combine  the luminosity distance measured by the supernovae
with the angular diameter distance measured by CMB.
The angular diameter distance
at recombination $z_*$, $D_A(z_*)$
is given by fitting the CMB acoustic structures.
We can estimate $D_A(z_*)$ based upon the cosmological parameter
$\omega_m=\Omega_m h^2= 0.135\pm 0.008$ which controls the physical extent of
the sound horizon as well as the redshift of recombination and
its measured angular scale $l_A=301\pm 1$ \cite{spergel03,peiris03}.

The right panel in Fig~\ref{fig:fitting} and 
the bottom left panel in Fig~\ref{fig:mu_w}
show that the p-p coupling scenario with no cosmic acceleration is now
more likely than either phantom models or
 $\Lambda$CDM at $w=-1$ but the latter is still allowed at
95\% confidence. 

A much more direct test of p-p dimming comes from angular diameter distance
measures to a redshift more comparable to the supernovae.  
This is provided by the recent detection of baryon oscillations in the SDSS
LRG galaxy survey.  This measurement can be summarized as 
a distance ratio  
\be
R \equiv { D_A(z=0.35) \over D_A(z_{*})} =0.0979\pm 0.0036 \,.
\ee
When this constraint is added in, a non accelerating universe with 
$w\ge -1/3$ is ruled out at the 4 $\sigma$ level (see  Fig~\ref{fig:fitting}).
Interestingly,  Fig.~\ref{fig:mu_w} shows two separate equally favored
solutions: one with negligible dimming and the other with
strong dimming.   

The favored p-p coupling with the current dataset requires
a strong external magnetic field with $\mu\sqrt{f} \sim 8 \times 10^{-30}$
which saturates the conversion probability
early in redshift.
As shown in the middle panel in Fig~\ref{fig:chromaticdimming},
the conversion probability with the strong external magnetic field
passes the chromaticity constraint test.

\section{conclusion}
We study an alternative explanation for the dimming of supernovae involving the loss of photons to 
a pseudo-scalar field $\phi$ in transit.
With only supernovae measurements on the bolometric total dimming and the achromaticity
of dimming, solutions without cosmic acceleration exist.   

This ambiguity in the 
supernovae data can be resolved with angular diameter distance measures that are
based on a standard length scale instead of a standard candle.    We apply the
CMB acoustic peaks and the recently detected baryon oscillations as constraints on the
model.  We also use CMB spectral constraints to eliminate a class of small magnetic field 
coherence length solutions.   We show that the non-accelerating solutions persist when adding just the
CMB information but are eliminated at the 4$\sigma$ level once the baryon oscillations are
added.  

Intriguingly solutions with substantial dimming accompanied by cosmic acceleration
still exist and could potentially bias measurements of the equation of state in the 
negative direction.   
These models can be tested with future baryon oscillation experiments.
Likewise astrophysical sources of non-cosmological dimming such as dust extinction
can be tested by these means.

\vspace{1.cm}

\noindent {\it Acknowledgments}:
YS and WH are supported by the   U.S.~Dept. of Energy contract DE-FG02-90ER-40560.
WH is additionally supported by the Packard Foundation.  This work was carried out at the
KICP under NSF PHY-0114422.

\appendix
\section{CMB spectral constraint}
\begin{figure}[htbp]
  \begin{center}
    \plotone{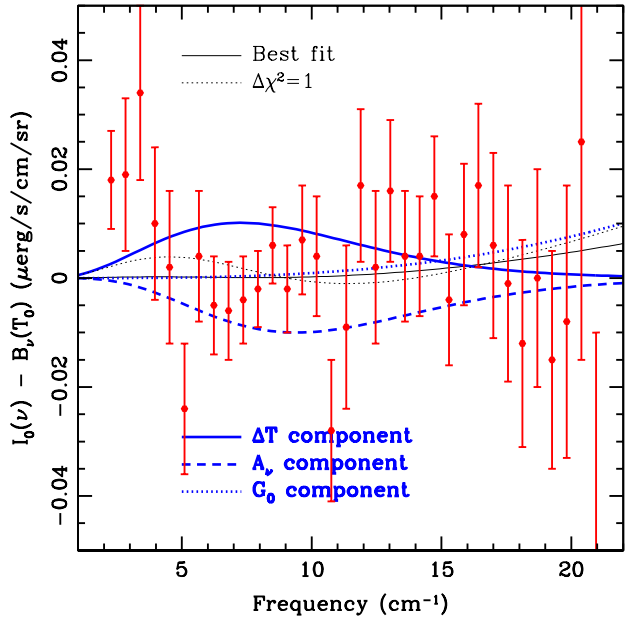}
    \caption{Residuals $I_0(\nu)-B_{\nu}(T_0)$ compared with model distortions. 
    Thin solid line denotes 
the best fit, thin dotted line denotes a deviation at the $\Delta \chi^2=1$ level.
This deviation is composed of compensating changes from the temperature variation
$\Delta T$, the p-p dimming $A_\nu$, and the galactic component $G_0$ as shown.
}
\label{fig:dblack}
\end{center}
\end{figure}

COBE FIRAS measured the CMB to be a pure blackbody with no significant distortions \cite{mather94}. 
A dimming of cosmological sources would distort the blackbody spectrum and so in this 
Appendix we use the COBE FIRAS data shown in Fig.~\ref{fig:dblack}
 to set limits on dimming scenarios.

The p-p coupling dims the CMB spectrum as well as the supernovae light.
We introduce the generic parameter $A_{\nu}$ representing the amplitude  of the
dimming of the CMB spectrum as
\ba
S(\nu;T)=\left[1-A_{\nu}\left(\frac{\nu}{10 {\rm cm}^{-1}}\right)^2\right]
\frac{2hc^2\nu^3}{\exp (hc\nu/kT)-1}\nn
\ea
where the frequency $\nu$ is measured in cm$^{-1}$.
Our linearized fit includes three unknown parameters, 
$\Delta T$ the temperature variation from $T_0=2.726 {\rm K}$, $A_{\nu}$ and
the level of residual galaxy emission $G_0$. The form is
\ba
I_0(\nu)=\left[1-A_{\nu}\left(\frac{\nu}{10}\right)^2\right]B_{\nu}(T_0)
+\Delta T \frac{\partial B_{\nu}}{\partial T}
+G_0g(\nu),,\nn
\ea
where $B_{\nu}(T_0)$ is the blackbody spectrum and $g(\nu)$ is a galaxy
emission spectrum.
We assume a  fixed galaxy emission spectrum over the sky as given in  \cite{wright94}.

As shown in Fig.~\ref{fig:dblack}, the best fit model is 
a pure blackbody spectrum of $T_{0}=2.726{\rm K}$
with a small high frequency galaxy emission contamination.
However the shape of the linear distortion by a temperature uncertainty
 $\Delta T$ is mildly degenerate with $A_\nu$.
Constraints on $A_{\nu}$ are limited by this degeneracy as shown in Fig.~\ref{fig:dblack}.
We obtain $A_{\nu}=1.0 \times 10^{-5}\pm 2.5\times 10^{-4}$
at the 95$\%$ confidence level.

\vfill


\end{document}